\documentclass[]{article}

\usepackage{graphicx}
\usepackage{amssymb}

\def\title#1{\begin{centering}\Large\bf #1 \\[5mm]\end{centering}}
\def\author#1{\begin{centering}\bf #1\end{centering}}
\def\address#1{\begin{centering}\bf #1 \\[5mm]\end{centering}}

\begin{document}
\small
\title{Expansion of a lithium gas in the BEC-BCS crossover}
\author{J.\,Zhang$^{a,b}$, E.\,G.\,M.\,
van Kempen$^c$, T.\,Bourdel$^a$, L.\,Khaykovich$^{a,d}$,
J.\,Cubizolles$^a$, F.\,Chevy$^a$\\ M.\,Teichmann$^a$
L.\,Tarruell$^a$, S.\,J.\,J.\,M.\,F.\,Kokkelmans$^{a,c}$, and
C.\,Salomon$^a$}

\address{$^a$ Laboratoire Kastler-Brossel, ENS,
24 rue Lhomond, 75005 Paris,\\$^b$ SKLQOQOD, Institute of
Opto-Electronics, Shanxi University, Taiyuan 030006, P.R.
China,\\$^c$ Eindhoven University of Technology, P.O.~Box~513,
5600~MB Eindhoven, The Netherlands,\\$^d$ Department of Physics,
Bar Ilan University, Ramat Gan 52900, Israel.}

\vskip10.pt

\begin{abstract}
We report on experiments in $^6$Li Fermi gases near Feshbach
resonances. A broad $s$-wave resonance is used to form a
Bose-Einstein condensate of weakly bound $^6$Li$_2$ molecules in a
crossed optical trap. The measured molecule-molecule scattering
length of $170^{+100}_{-60}$\,nm at 770\,G is found in good
agreement with theory. The expansion energy of the cloud in the
BEC-BCS crossover region is measured. Finally we discuss the
properties of $p$-wave Feshbach resonances observed near
200\,Gauss and new $s$-wave resonances in the heteronuclear
$^6$Li- $^7$Li mixture.
\end{abstract}

Strongly interacting fermionic systems occur in a variety of
physical processes, ranging from nuclear physics, to high
temperature superconductivity, superfluidity, quark-gluon plasmas,
and ultra-cold dilute gases. Thanks to the phenomenon of Feshbach
resonances, these gases offer the unique possibility  to tune the
strength and the sign of the effective interaction between
particles. In this way, it is possible to study the crossover
between situations governed by Bose-Einstein and Fermi-Dirac
statistics.
\section{BEC-BCS crossover near $^6$Li s-wave resonance}

 When the scattering length $a$ characterizing
the 2-body interaction at low temperature is positive, the atoms
can pair in a weakly bound molecular state. When the temperature
is low enough, these bosonic dimers can form a Bose-Einstein
condensate (BEC) as observed very recently in $^{40}$K
\cite{Greiner03} and $^6$Li \cite{Jochim03,Zwierlein03,Bourdel04}.
On the side of the resonance where $a$ is negative, one expects
the well known Bardeen-Cooper-Schrieffer (BCS) model for
superconductivity to be valid. However, this  simple picture of a
BEC phase on one side of the resonance and a BCS phase on the
other is valid only for small atom density $n$. When
$n|a|^3\gtrsim 1$ the system enters a strongly interacting regime
that represents a challenge for many-body theories
\cite{Leggett80}. In the recent months, this regime has been the
subject of intense experimental activity
\cite{Bourdel04,Bartenstein04,Regal04,Zwierlein04,Greiner04,Kinast04,Chin04}.

Here we first report on Bose-Einstein condensation of $^6$Li
dimers in a crossed optical dipole trap, and a study of the
BEC-BCS crossover region. Unlike all previous observations of
molecular BEC made in single beam dipole traps with very elongated
geometries, our condensates are formed in nearly isotropic
strongly confining traps. The experimental setup has been
described previously \cite{Bourdel03,Cubizolles03}. A gas of
$^6$Li atoms is prepared in the absolute ground state
$|1/2,1/2\rangle$ in a Nd-YAG crossed beam optical dipole trap.
The horizontal beam (resp.\,vertical) propagates along $x$ ($y$),
has a maximum power of $P_o^h=2\,$W ($P_o^v=3.3\,$W) and a waist
of $\sim 25\,\mu$m ($\sim 40\,\mu$m). At full power, the $^6$Li
trap oscillation frequencies are $\omega_x/2\pi= 2.4(2)\,$kHz,
$\omega_y/2\pi=5.0(3)\,$kHz, and $\omega_z/2\pi=5.5(4)$\,kHz, as
measured by parametric excitation, and the trap depth is $\sim
80\,\mu$K. After sweeping the magnetic field $B$ from 5\,G to
1060\,G, we drive the Zeeman transition between $|1/2,1/2\rangle$
and $|1/2,-1/2\rangle$ with a 76\,MHz RF field to prepare a
balanced mixture of the two states. As measured recently
\cite{Chin04}, the Feshbach resonance between these two states is
peaked at 834(2)\,G, and for $B$=1060\,G, $a=-167$\,nm. After
100\,ms the coherence between the two states is lost and plain
evaporation provides $N_\uparrow=N_\downarrow=N_{\rm
tot}/2=1.5\times 10^5$ atoms at 10\,$\mu$K=0.8\,$T_{\rm F}$, where
$k_{\rm B}T_{\rm F}=\hbar^2k_{\rm F}^2/2m=\hbar(3 N_{\rm
tot}\omega_x \omega_y \omega_z)^{1/3}=\hbar\bar{\omega}(3 N_{\rm
tot})^{1/3}$ is the Fermi energy. Lowering the intensity of the
trapping laser to $0.1\,P_0$, the Fermi gas is evaporatively
cooled to temperatures $T$ at or below $0.2\,T_{\rm F }$ and
$N_{\rm tot}\approx 7\times 10^4$.

\begin{figure}[hb]
\begin{center}
\includegraphics[width=11cm]{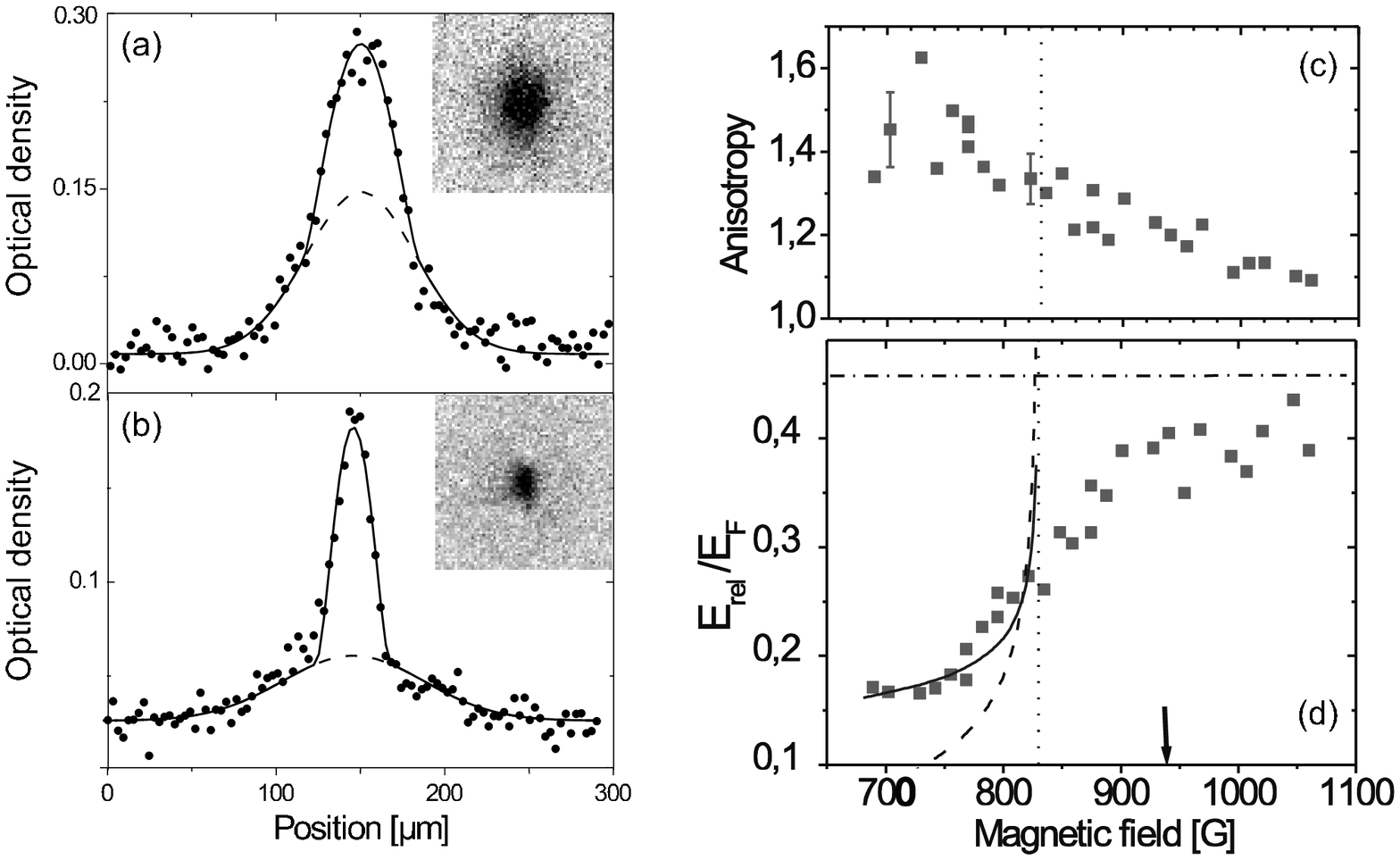}
\caption{\label{fig:figure1} \small a,b: Onset of Bose-Einstein
condensation in a cloud of $2\times 10^4$ $^6$Li dimers at 770\,G
(a) and of $2\times 10^4$ $^7$Li atoms at 610\,G (b) in the same
optical trap. (a): 1.2\,ms expansion profiles along the weak
direction $x$ of confinement. (b): 1.4 ms expansion. The different
sizes of the condensates reflect the large difference in
scattering length $a_{\rm m}=170\,$nm for $^6$Li dimers and
$a_{7}=0.55$\,nm for $^7$Li. Solid line: Gaussian+Thomas-Fermi
fit. Dashed line: gaussian component. Condensate fractions are
$44\,\%$ in (a) and $28\%$ in (b). c,d: BEC-BCS crossover. (c):
anisotropy of the cloud. (d): release energy across the BEC-BCS
crossover region. In (d), the dot-dashed line corresponds to a
$T=0$ ideal Fermi gas. The dashed curve is the release energy from
a pure condensate in the Thomas-Fermi limit. The solid curve
corresponds to a finite temperature mean field model with
$T=0.5\,T_{\rm C}^0$. Arrow: $k_{\rm F}|a|=3$.}
\end{center}
\end{figure}
Then, sweeping  the magnetic field to $770\,$G in $200\,$ms, the
Feshbach resonance is slowly crossed. In this process atoms are
adiabatically and reversibly transformed into cold molecules
\cite{Cubizolles03,Regal03} near the BEC critical temperature as
presented in figure 1a.  The onset of condensation is revealed by
bimodal and anisotropic momentum distributions in time of flight
expansions of the molecular gas. These images are recorded as
follows. At a fixed magnetic field, the optical trap is first
switched off. The cloud expands typically for 1\,ms and then the
magnetic field is increased by 100 \,G in $50\,\mu$s. This
converts the molecules back into free atoms above resonance
without releasing their binding energy \cite{Zwierlein03}.
Switching the field abruptly off in $10\,\mu$s, we detect free
$^6$Li atoms by light absorption near the D2 line. We have checked
that, in the trap before expansion, there are no unpaired atoms.
In figure 1b, a Bose-Einstein condensate of $^7$Li atoms produced
in the same optical trap is presented. The comparison between the
condensate sizes after expansion dramatically reveals that the
mean field interaction and scattering length are much larger for
$^6$Li$_2$ dimers (Fig.\,1a) than for $^7$Li atoms (Fig.\,1b).

To measure the molecule-molecule scattering length, we produce
pure molecular condensates by taking advantage of our crossed
dipole trap. We recompress the horizontal beam to full power while
keeping the vertical beam at the low power of 0.035\,$P_0^v$
corresponding to a trap depth for molecules $U=5.6\,\mu$K.
Temperature is then limited to $T\leq 0.9\,\mu$K assuming a
conservative $\eta=U/k_{\rm B}T=6$, whereas the critical
temperature increases with the mean oscillation frequency.
Consequently, with an axial (resp. radial) trap frequency of
440\,Hz (resp. 5\,kHz), we obtain $T/T_{\rm C}^0\leq 0.3$, where
$T_{\rm C}^0=\hbar\bar{\omega}(0.82N_{\rm
tot}/2)^{1/3}=$2.7\,$\mu$K is the non interacting BEC critical
temperature. Thus, the condensate should be pure as confirmed by
our images.
 After 1.2~ms of expansion, the radius of the
condensate in the $x$  (resp. $y$) direction is $R_x=51~\mu$m
($R_y=103~\mu$m). The resulting anisotropy $R_y/R_x=2.0(1)$ is
consistent with the value 1.98 \cite{correctioncourbure} predicted
the scaling equations \cite{Kagan96,Castin96}. Moreover, this set
of equation  leads to an {\em in-trap} radius $R_x^0=26\mu$m
(resp. $R_y^0=2.75\mu$m). We then deduce the molecule-molecule
scattering length from the Thomas-Fermi formula $R_{x,y}^0= a_{\rm
ho}\bar\omega/\omega_{x,y}(15 N_{\rm tot}a_{\rm m}/2a_{\rm
ho})^{1/5}$, with $a_{\rm ho}=\sqrt{\hbar/2m\bar\omega}$.
Averaging over several images, this yields $a_{\rm
m}=$170$^{+100}_{-60}$\,nm at 770\,G. Here, the statistical
uncertainty is negligible compared to  the systematic uncertainty
due to the calibration of our atom number.  At this field, we
calculate an atomic scattering length of $a= 306$\,nm. Combined
with the prediction $a_{\rm m}=0.6\,a$ of \cite{Petrov04},  we
obtain $a_{\rm m}=183\,$nm in good agreement with our measurement.
For $^7$Li, we obtain with the same analysis a much smaller
scattering length of $a_7$=0.65(10)\,nm at 610\,G also in
agreement with theory \cite{Khaykovich02}.

The condensate lifetime is typically $\sim$300\,ms at 715\,G
($a_{\rm m}=66\,$nm) and $\sim$3\,s at 770\,G ($a_{\rm
m}=170\,$nm), whereas for $a=-167$\, nm at 1060\,G, the lifetime
exceeds 30\,s. On the BEC side, the molecule-molecule loss rate
constant is $G=0.26^{+0.08}_{-0.06} \times 10^{-13}\,$cm$^3$/s at
770\,G and $G=1.75^{+0.5}_{-0.4} \times 10^{-13}\,$cm$^3$/s at
715\,G with the fit procedure for condensates described in
\cite{Soding97}.  Combining similar results for four values of the
magnetic field ranging from 700~G to 770~G, we find $G\propto
a^{-1.9 \pm 0.8}$ (figure 2). Our data are in agreement with the
theoretical prediction $G \propto a^{-2.55}$ of
ref.\,\cite{Petrov04} and with
 previous measurements of G in a thermal gas at 690\,G
\cite{Cubizolles03} or in a BEC at 764~G \cite{Bartenstein04}. A
similar power law was also found for $^{40}$K \cite{Regal04b}.

\begin{figure}[ht]
\begin{center}
\includegraphics[width=8cm]{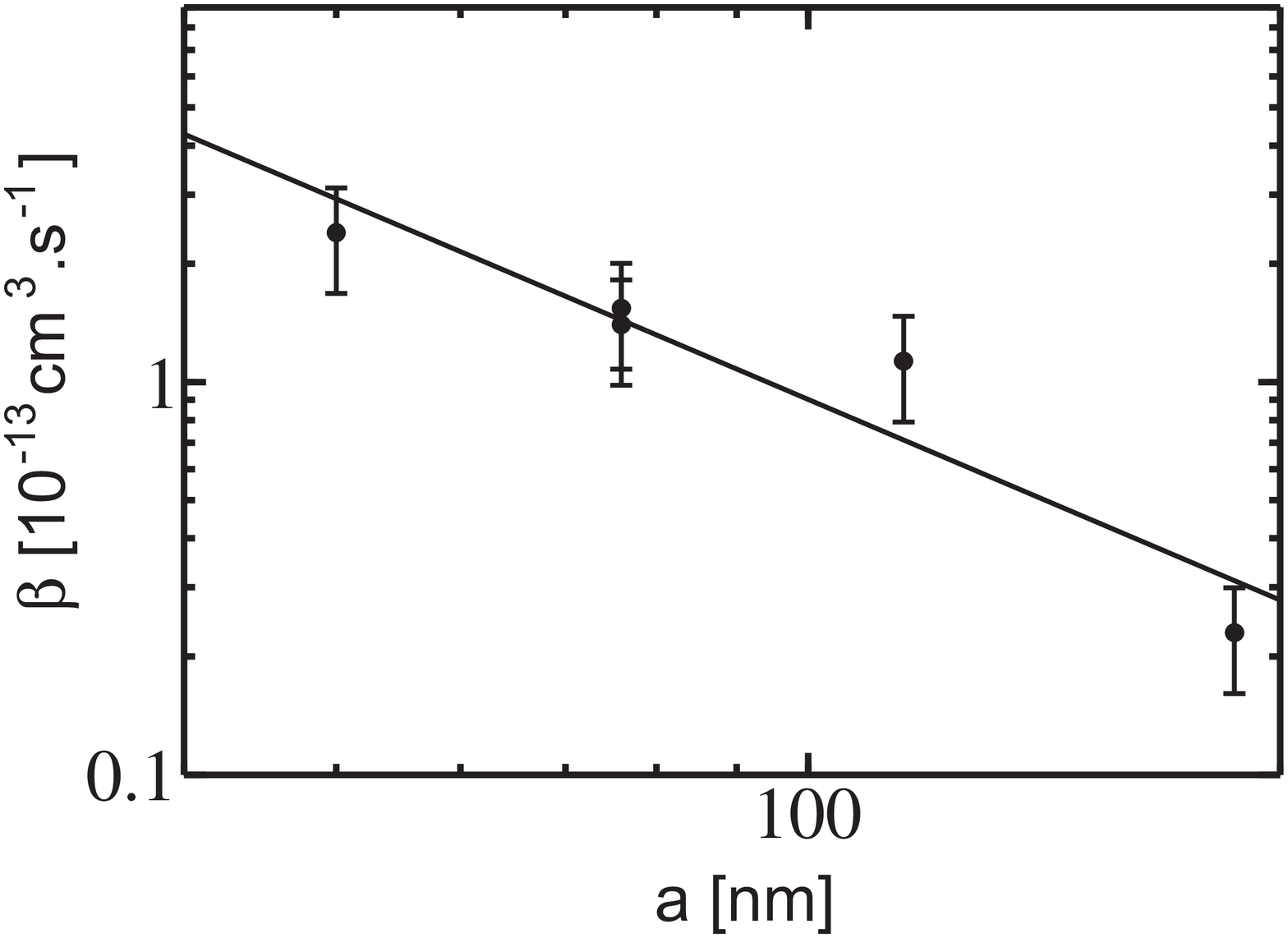}
\caption{\label{fig:figure2} \small Molecular condensate loss rate
$\beta$ as a function of the atomic scattering length $a$ near the
834 G s-wave Feshbach resonance. The line is a power law fit with
$\beta \sim a^{-1.9\pm 0.8}$ in agreement with theory, $\beta \sim
a^{-2.55}$ \cite{Petrov04}.}
\end{center}
\end{figure}

We then made an investigation of the crossover from a
Bose-Einstein condensate to an interacting Fermi gas (Fig.
\ref{fig:figure1}.c and d). We prepare a nearly pure condensate
with 3.5$\times 10^4$ molecules at 770\,G and recompress the trap
to frequencies of $\omega_x=2 \pi \times 830\,$Hz, $\omega_y=2 \pi
\times 2.4\,$kHz, and $\omega_z=2 \pi \times 2.5\,$kHz. The
magnetic field is then slowly swept at a rate of 2\,G/ms to
various values across the Feshbach resonance. The 2D momentum
distribution after a time of flight expansion of 1.4\,ms is then
detected as previously. As $B$ increases from the regime of weak
interactions the condensate size gradually increases towards the
width of a non interacting Fermi gas. Nothing particular happens
on resonance. Fig. \ref{fig:figure1}.c and \ref{fig:figure1}.d
present respectively the anisotropy of the cloud after expansion
$\eta$ and the corresponding released energy  $E_{\rm rel}$. These
are calculated from gaussian fits to the density after time of
flight: $E_{\rm rel}=m (2\sigma_y^2+\sigma_x^2)/2\tau^2$ and
$\eta=\sigma_y/\sigma_x$, where $\sigma_i$ is the rms width along
$i$, and $\tau$ is the time of flight. The anisotropy
monotonically decreases from $\sim$1.6 on the BEC side, where
hydrodynamic expansion predicts 1.75, to 1.1, at 1060\,G, on the
BCS side. On resonance, at zero temperature, a superfluid
hydrodynamic expansion is expected \cite{Menotti02} and would
correspond to $\eta=1.7$.  We find however $\eta=1.35(5)$,
indicating a partially hydrodynamic behavior that could be due to
a reduced superfluid fraction. On the $a<0$ side, the decreasing
anisotropy would indicate a further decrease of the superfluid
fraction that could correspond to the reduction of the condensed
fraction of fermionic atom pairs away from resonance observed in
\cite{Regal04,Zwierlein04}. Interestingly, our results  differ
from that of ref.\cite{OHara02} where hydrodynamic expansion was
observed at $910\,$G in a more elongated trap for $T/T_{\rm
F}\simeq 0.1$.

In the BEC-BCS crossover regime, the gas energy released after
expansion $E_{\rm rel}$ is also smooth (Fig. \ref{fig:figure1}.d).
$E_{\rm rel}$ presents a plateau for $B\leq 750\,$G, and then
increases monotonically towards that of a weakly interacting Fermi
gas. The plateau is not reproduced by the mean field approach of a
pure condensate (dashed line). This is a signature that the gas is
not at $T=0$. It can be understood with the mean field approach we
used previously to describe the behavior of the thermal cloud.
Since the magnetic field sweep is slow compared to the gas
collision rate \cite{Cubizolles03}, we assume that this sweep is
adiabatic and conserves entropy \cite{Carr03}. We then adjust this
entropy to reproduce the release energy at a particular magnetic
field, $B=720$\,G. The resulting curve as a function of $B$ (solid
line in Fig.\,\ref{fig:figure1}.d) agrees well with our data in
the range $680\,$G$\,\leq B\leq 770\,$G, where the condensate
fraction is 40$\%$, and the temperature is $T\approx T_{\rm
C}^0/2=\,1.4\,\mu$K. This model is limited to $n_ma_{\rm m}^3
\lesssim 1$. Near resonance the calculated release energy diverges
and clearly departs from the data. On the BCS side, the release
energy of a $T=0$ ideal Fermi gas gives an upper bound for the
data (dot-dashed curve), as expected from negative interaction
energy and a very cold sample. This low temperature is supported
by our measurements on the BEC side and the assumption of entropy
conservation through resonance which predicts $T=0.1\,T_{\rm F}$
on the BCS side \cite{Carr03}.

 On resonance the gas is expected to reach a universal
behavior, as the scattering length $a$ is not a relevant parameter
any more \cite{Heiselberg01}. In this regime, the release energy
scales as $E_{\rm rel}=\sqrt{1+\beta} E_{\rm rel}^0$, where
$E_{\rm rel}^0$ is the release energy of the non-interacting gas
and $\beta$ is a universal parameter. From our data at 820~G, we
get $\beta=-0.64(15)$. This value is larger than the Duke result
$\beta=-0.26\pm 0.07$ at 910\,G \cite{OHara02}, but agrees with
that of Innsbruck $\beta=-0.68^{+0.13}_{-0.10}$ at 850\,G
\cite{Bartenstein04}, and with the most recent theoretical
prediction $\beta=-0.56$ \cite{Carlson03,Astrakharchik04}. Around
925\,G, where $a=-270\,$nm and $(k_{\rm F}|a|)^{-1}=0.35$, the
release energy curve displays a change of slope. This is a
signature of the transition between the strongly and weakly
interacting regimes. It is also observed near the same field in
\cite {Bartenstein04} through {\it in situ} measurement of the
trapped cloud size. Interestingly, the onset of resonance
condensation of fermionic atom pairs observed in $^{40}$K
\cite{Regal04} and $^{6}$Li \cite{Zwierlein04}, corresponds to a
similar value of $k_{\rm F}|a|$.

\section{P-wave resonances}

Recently, both experimental \cite{JinPWave,Zhang,Ketterle} and
theoretical \cite{Ho} papers have devoted interest to the
 p-wave Feshbach resonances. The goal of these
experiments is to nucleate molecules with internal angular
momentum $l=1$, that could lead to the observation of some
non-conventional superconductivity, analogous to that observed in
superfluid $^3$He \cite{RefHe3}.

\begin{figure}[h]
\includegraphics[width=\columnwidth]{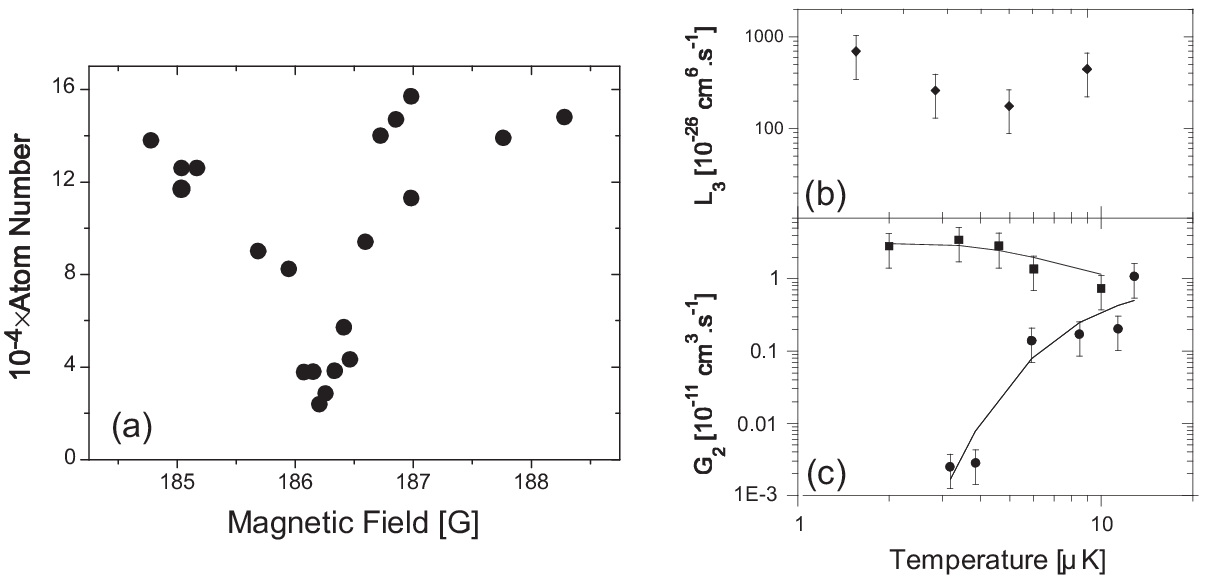}
\caption{\small (a) Atom number vs. magnetic field $B_{0,{\rm f}}$
after a 50~ms wait for atoms  in the spin mixture $(1/2,-1/2)$ at
$T\sim 14\mu$K. The sharp drop close to $B_0\sim 186$~G over a
range $\simeq 0.5$~G is the signature of the p-wave Feshbach
resonance predicted by theory. (b) (resp. (c)) Variations of
3-body (2-body)  loss rates vs temperature at the Feshbach
resonance. (b): $\blacklozenge$: atoms in the Zeeman ground state
$|f=1/2,m_f=1/2\rangle$, $B_{0,{\rm f}}\sim 159$~G. (c):
 $\blacksquare$: atoms polarized in
$|f=1/2,m_f=-1/2\rangle$, $B_{0,{\rm f}}\sim 215$~G. $\bullet$:
mixture $|f=1/2,m_f=1/2\rangle+|f=1/2,m_f=-1/2\rangle$, $B_{0,{\rm
f}}\sim 186$~G. In both cases, the full line is a fit to the data
using prediction of Eq. \ref{Eqn2Body} with the magnetic field as
the only fitting parameter. } \label{Fig3}
\end{figure}

In the manifold $f=1/2$ corresponding to the hyperfine ground
state of $^6$Li, coupled channels calculations have demonstrated
that three  Feshbach resonances could be observed in p-wave
channels. The position of the resonances are  calculated using the
most recent experimental data available on $^6$Li and are
presented in Tab. \ref{table1}. The predicted values of the
resonance position are compared with the location we obtained
experimentally using the following procedure: we prepare  $^6$Li
atoms in the dipole trap with the required spin state $(m_f,m_f')$
using radiofrequency transfer. We then ramp up the magnetic field
from 0 to a value $B_1$ slightly higher than the predicted
position of the Feshbach resonance. The magnetic field is then
abruptly decreased to a variable value $B$ close to resonance.
After a waiting time of 50\,ms in the trap, we measure the atom
number by time of flight imaging. In Fig. \ref{Fig3}.a, we show
the evolution of the atom number vs magnetic field in the channel
(1/2,-1/2). We observe a sharp decrease of the atom number at
$\sim 186.5$\, G, close to the predicted value 185\, G. The two
other channels display similar losses in the vicinity of the
theoretical position of the Feshbach resonances (Tab.
\ref{table1}). Note that in this table the experimental
uncertainty is mainly due to the magnetic field calibration.

One of the main issue of the physics of Feshbach resonances is
related to the lifetime of the molecules, and more generally of
the atoms, at resonance. Indeed, one of the key element that led
to the experiments on the BEC-BCS crossover was the increase of
the lifetime of molecules composed of fermions close to resonance.
To address this issue, we have measured the time evolution of the
atom number $N$ in the sample at the three Feshbach resonances.
Since the one-body lifetime is $\sim 30$\, s,  much longer than
the measured decay time ($\sim 100$\, ms), we can fit the time
evolution using the rate equation

\begin{equation} \frac{\dot N}{N}=-G_2\langle n\rangle-L_3\langle
n^2\rangle, \label{RateEqn}
\end{equation}

\noindent where $n$ is the atom density and $\langle
n^a\rangle=\int {\rm d}^3r\, n^{a+1}/N$ ($a=1,2$) is calculated
from the classical Boltzman distribution.

In contrast to s-wave Feshbach resonances where dipolar losses are
forbidden in the $f=1/2$ manifold \cite{Dieckmann02},  the losses
near a p-wave resonance are found to be dominantly 2-body in the
(1/2,-1/2) and (-1/2,-1/2) channels.  The variations of the 2-body
loss rate with temperature are displayed in Fig. \ref{Fig3}.c and
show very different behaviors in these two channels. This non
trivial dependence is actually not the consequence of some
specific property of the states involved (eg., the quantum
statistics) but can be recovered using a very simple three-state
model.  We describe inelastic processes by two non interacting
open channels, respectively the incoming and decay channels, that
are coupled to a single p-wave molecular state . This model leads
to a very simple algebra that can be treated analytically
\cite{FredPWave} and yields for the two-body loss-rate at a given
energy $E$

\begin{equation}
g_2 (E)=\frac{K E}{(E-\delta)^2+\gamma^2/4} \label{Eqn2BodyA}.
\end{equation}

\noindent Here $\delta=\mu (B-B_{\rm F})$ is the detuning to the
Feshbach resonance and $K$, $\mu$ and $\gamma$ are
phenomenological constants depending on the microscopic details of
the potential. For each channel, these parameters are estimated
from our numerical coupled-channel calculation  (Tab.
\ref{table1}).

Eqn. (\ref{Eqn2BodyA}) shows that, contrarily to s-wave processes
that can be described accurately by their low energy behavior,
p-wave losses are dominated by the resonance peak located at
$E=\delta$ where the losses are maximum. In other word, the
so-called ``threshold laws", that give the low energy scattering
behavior, are insufficient to describe  a Feshbach resonance
associated with p-wave molecular states (and, more generally, any
non zero angular momentum molecular state).

\begin{table}
\centerline{\begin{tabular}{cccccc} \hline\hline
$(m_{f_1},m_{f_2})$&
\begin{tabular}{c}
$B_{\rm th}$\\
G
\end{tabular}&
\begin{tabular}{c}
$B_{\rm exp}$\\
G
\end{tabular}&
\begin{tabular}{c}
$K$\\
${\rm cm^3\cdot \mu K\cdot s^{-1}}$
\end{tabular}&
\begin{tabular}{c}
$\gamma$\\
${\rm \mu K}$
\end{tabular}&
\begin{tabular}{c}
$\mu$\\
${\rm \mu K\cdot G^{-1}}$
\end{tabular}
\\
\hline
(1/2,1/2) & 159 & 160.2(6)& -- & -- & -- \\
(1/2,-1/2)&185 & 186.2(6)&$1.21\times 10^{-13}$&0.05&117\\
(-1/2,-1/2)& 215 & 215.2(6)&$7.33\times 10^{-13}$&0.08&111\\
\hline\hline
\end{tabular}}
\caption{\label{table1} Position and two-body losses parameters of
the p-wave Feshbach resonances of $^6$Li atoms in
$|f_1=1/2,m_{f_1}\rangle$ and $|f_2=1/2,m_{f_2}\rangle$
 .}
\end{table}

To compare with experimental data, Eq. (\ref{Eqn2BodyA}) is
averaged over a thermal distribution and for $\delta>0$ and
$\delta\gg\gamma$ we get:

\begin{equation}G_{2}\sim
4\sqrt{\pi}\frac{K}{\gamma}\left(\frac{\delta}{k_{\rm B}
T}\right)^{3/2}{\rm e}^{-\delta/k_{\rm B} T}. \label{Eqn2Body}
\end{equation}

Eqn. \ref{Eqn2Body} is used to fit the data of Fig. \ref{Fig3}.b,
with $B-B_{\rm F}$ as the only fitting parameter. We get a fairly
good agreement if we take $B-B_{\rm F}=0.04$~G (resp. $0.3$~G) for
the (-1/2,-1/2) (resp. (1/2,-1/2)) channel, illustrating the
extreme sensitivity of $G_2$ to detuning and temperature. This
feature was also tested by measuring the variations of $G_2$ with
magnetic field at constant temperature. Note that the width of the
resonance, as given by Eqn. \ref{Eqn2Body}, is of the order of
$k_{\rm B} T/\mu$. At $\sim 10\, \mu$K, this yields a width of
$\sim 0.1$\, G, which is comparable with the one observed in Fig.
\ref{Fig3}.a.

In the (1/2,1/2) channel, dipolar losses are forbidden and we
indeed find that 3-body losses are dominant.
 The dependence of $L_3$ with temperature is very weak  (Fig.
\ref{Fig3}.b) and contradicts the low energy theoretical
calculation of the temperature dependence of three-body loss rate
performed in \cite{Esry02}. Indeed, Wigner threshold law predicts
that at low energy, $L_3$ should be proportional to $T^\lambda$,
with $\lambda\ge 2$  for indistinguishable fermions.  However, as
we noticed earlier in the case of two-body losses, the threshold
law is probably not sufficient to describe losses at resonance due
to the existence of a resonance peak that might also be present in
three-body processes. This suggests that 3-body processes must be
described by a more refined formalism,
 analogous to the unitary limited treatment of the s-wave
elastic collisions \cite{Esry03}.

 Finally, we have
checked the production of molecules in (1/2,-1/2)mixture by using
the scheme presented in \cite{Cubizolles03,Regal03}. We first
generate $p$-wave molecules  by ramping in 20~ms the magnetic
field from 190~G$>B_{\rm F}$ to $B_{\rm nuc}=185~{\rm G}<B_{\rm
F}$. At this stage, we can follow two paths before detection. Path
1 permits to measure the number $N_1$ of free atoms: by ramping
{\it down} in 2~ms the magnetic field from 185~G to 176~G, we
convert the molecules into deeply bound molecular states that
decay rapidly by 2-body collisions. Path 2 gives access to the
total atom number $N_2$ (free atoms + atoms bound in p-wave
molecules). It consists in ramping {\it up} the magnetic field in
2~ms from $B_{\rm nuc}$ to 202~G$>B_{\rm F}$ to convert the
molecules back into atoms. Since the atoms involved in molecular
states appear only in pictures taken in path 2, the number of
molecules in the trap is $(N_2-N_1)/2$.  In practice, both
sequences are started immediately after reaching $B_{\rm nuc}$ and
we average the data of 25 pictures to compensate for atom number
fluctuations. We then get $N_1=7.1(5)\times 10^4$ and
$N_2=9.1(7)\times 10^4$ which corresponds to a molecule fraction
$1-N_1/N_2=0.2(1)$. Surprisingly, we failed to detect any molecule
signal  when applying the same method to (1/2,1/2) atoms.

\section{Heteronuclear Feshbach resonances}

So far, the Feshbach resonances used in this paper were involving
atoms of the same species (namely $^6$Li). However, it was
recently pointed out that the observation of Feshbach resonances
between two different atom species could lead to a host of
interesting effects ranging from the observation of supersolid
order \cite{Buchler03} to the study polar molecules
 \cite{Baranov03}. In the case of a mixture of bosons and
fermions, these molecules are fermions and are expected to be long
lived. Indeed,  Pauli principle keeps molecules far apart and
prevents inelastic collisions. Such resonances were observed
experimentally in $^6$Li-$^{23}$Na \cite{Stan04} and
$^{40}$K-$^{87}$Rb \cite{Inouye04} mixtures. In the case of
$^6$Li-$^7$Li  in the stable Zeeman ground state $
|f=1/2,m_f=1/2\rangle\otimes |f=1,m_f=1\rangle$, the existence and
the position of heteronuclear Feshbach resonances were predicted
in \cite{VanKempen04}. In that work, the $^6$Li-$^7$Li interaction
potential was extracted from the data on $^6$Li-$^6$Li scattering
properties by mean of a simple mass-scaling. Using coupled
channels calculation it was found that this system exhibited five
Feshbach resonances whose position is displayed in Tab.
\ref{TableHetero} \cite{ResonanceNumber}.

Experimentally, we probed these Feshbach resonances using a
mixture of respectively $N_6\sim N_7\sim 10^5$ atoms of $^6$Li and
$^7$Li in the absolute Zeeman ground state $|1/2,1/2\rangle\otimes
|1,1\rangle$. The gas is cooled at a temperature of $4\,\mu$K in
the cross dipole trap and  we locate the resonance  by sweeping
down in 1\,s the magnetic field from a value located about 20~G
above the predicted position of the resonance to a variable value
$B$. By looking for the value of $B$ at which we start loosing
atoms, we were able to detect four of the five resonances at a
value very close to that predicted by theory (Tab.
\ref{TableHetero}). The discrepancy between the experimental and
theoretical values is larger in this case than in the case of th
p-wave resonance. This is probably due to a breakdown of the
Born-Oppenheimer approximation that one expects in the case of
light atoms such as lithium and that forbids the use of the mass
scaling \cite{VanKempen04}. Note also that the missing resonance
is predicted to be very narrow ($\sim$1\, mG).

As noticed in \cite{Pethick}, the width of a Feshbach resonance
strongly influences the  molecule lifetime that can be much higher
close to a wide resonance. Since the Feshbach resonances we found
in $|1/2,1/2\rangle\otimes |1,1\rangle$ are very narrow ($\sim
0.1$\, G) it might prove interesting to look for wider Feshbach
resonances. In the case of $^6$Li-$^7$Li, a wide resonance is
predicted to exist in the stable state $|3/2,3/2\rangle\otimes
|1,1\rangle$ at $\sim 530$\, G, in agreement with our observation
of a large atom loss located between $\sim 440$\, G and $\sim
540$\, G.

\begin{table}[t]
\centerline{\begin{tabular}{cc}
\hline\hline$B_{\rm th}$ (G)&$B_{\rm exp}$ (G)\\
\hline
218&not seen\\
230&226.3(6)\,G\\
251&246.0(8)\,G\\
551&539.9(8)\,G\\
559&548.6(9)\,G\\
\hline\hline
\end{tabular}}
\caption{\label{TableHetero}\small Predicted and observed
positions of the heteronuclear Feshbach resonances of a
$^6$Li-$^7$Li gas. First column: resonance position as predicted
by coupled channel calculation. Secund column: experimental
determination of the resonances. The experimental error bars are
due to the systematic uncertainty on the calibration of the
magnetic field. The first Feshbach resonance is extremely narrow
and could not be detected with our experimental resolution.
 }
\end{table}


\begin{thebibliography}{99}
\bibitem{correctioncourbure} We correct our data for the presence of
a magnetic field curvature which leads to an anti-trapping
frequency of 100\,Hz at 800\,G along $x$.
\bibitem{ResonanceNumber} Note that the absence of symmetrization constraint in the case
of a heteronuclear system allows for many more resonances than for
a pure system.
\end{thebibliography}
\end{document}